\documentstyle[aps,prb,twocolumn,epsfig]{revtex}
\def\be{\begin{equation}}
\def\ee{\end{equation}}

\def\e{\varepsilon}

\begin{document}

\title{Electronic transport in strongly anisotropic disordered systems:
 model for the random matrix theory  with non-integer $\beta$ }

\author{Peter Marko\v s$^{*}$\\
Institute of Physics, Slovak Academy of Sciences, D\'ubravsk\'a cesta 9,
 842 28 Bratislava, Slovakia}
\maketitle

\begin{abstract}
We study numerically electronic transport in strongly 
anisotropic weakly disordered two-dimensional 
systems. We find that the  conductance 
distribution is Gaussian. The  conductance fluctuations
increase when  anisotropy becomes stronger.
The statistics  of the transport parameters can be interpreted by the
random matrix theory with 
a non-integer symmetry parameter $\beta$. Our results are 
in  a good agreement  with recent theoretical work
of K.A. Muttalib and J.R. Klauder [Phys. Rev. Lett. {\bf 82} (1999) 4272]
~~\\
\end{abstract}

It is generally accepted that
the electronic transport in weakly disordered metallic systems  is 
successfully described by the random matrix theory (RMT)
\cite{pichardnato}
and the Dorokhov-Mello-Pereyra-Kumar (DMPK) equation.
\cite{DMPK} Both theories predict a Gaussian distribution of the conductance
and provides us with  the  exact value of the conductance fluctuations in agreement
with data obtained by  diagrammatic expansion.
\cite{stone}
The Landauer formula for the conductance,
$g= {\rm Tr}~t^{\dag}t$
enables us to express $g$ in terms of  eigenvalues of the transmission matrix 
$t^\dag t$:
\be\label{g}
g=\sum_i^{N}\cosh^{-2} (z_i/2).
\ee
where $N$ is the number of open channels.

In the limit $N>>1$, 
RMT proposed a common probability distribution of parameters $z$ 
\be\label{rmtp}
P(z)=\exp -{\cal H}(z)/\beta
\ee
with
\be\label{rmt}
{\cal H}(z)=\sum_{i}^{N}\frac{k}{2}z_i^2-\sum_i^{N}J(z_i)-\beta\sum_{i<j}^N u(z_i,z_j).
\ee

In RMT, physical properties of the sample are specified  only by two parameters:
$\beta=1,2,4$ for
the orthogonal, unitary and symplectic symmetry of the model, respectively, 
and
the  mean free path $l$, which determines, together with the system length $L_x$,
the strength of one particle potential: 
$k=l N/L_x$.
RMT describes successfully the transport properties of 
weakly disordered  quasi-one dimensional (Q1D)
systems. It could be applied also to 
squares \cite{pichard}  or cubes  \cite{MKr} if the length $L_x$ of the system
fulfills the relations
\be\label{limits}
l<<L_x<\xi.
\ee
with $\xi$ being the localization length.
The absence of any other parameters reveals the universal transport properties of weakly disordered systems. In particular, the variance of the conductance is universal, 
and depends only on the symmetry and the shape of the sample. \cite{stone,melostone,RMS}

Recently, Muttalib and Klauder \cite{MK} showed 
that the  requirement of the large system length is not
necessary for the derivation of the DMPK equation. 
In their DMPK equation  
the parameter $\beta$  depends on the statistical properties of the model. This
allows $\beta$ to possess any positive value. As supposed, $\beta$ converges to unity
when the system length increases.

\medskip

In this paper we  present a physical realization of the theoretical model
proposed in Ref. \cite{MK}.
We calculate the conductance and the statistics of the parameters $z$ for weakly
disordered strongly anisotropic two-dimensional systems and show that their transport properties
can be described within the  framework of the RMT with the symmetry parameter $\beta$ smaller 
than 1.

Our model is defined by 
two dimensional (2D) anisotropic Anderson Hamiltonian
\be\label{ham}
H=\sum_{ij}\e_{ij}|ij\rangle\langle ij|
+\sum_i|ij\rangle\langle i+1j|+ t\sum_j|ij\rangle\langle ij+1| 
\ee
where $i\le L_x$ ($j\le L_y$) numerates sites in $x$ ($y$) direction, respectively.
Hard wall boundary conditions are considered and $E_F=0$. Then $N=L_y$.
Random energies $\e_{ij}$ are distributed uniformly between $-W/2$ and
$W/2$.
We put  $W=2$ throughout the paper. 
Then the localization length increased from $\xi\sim 25$ 
for the one dimensional (1D) chain ($t=0$) 
\cite{CK}
to  $\sim 10^6$ for the  2D isotropic systems
\cite{MacK} while the mean free path decreases from $l\sim 25$ to $l\sim 4$
~\cite{pichard}
in the same range of $t$. 
As a  typical size of our samples varies between 20-100, we expect to find
metallic behavior even for strong anisotropy.

\smallskip

We found that the conductance distribution is Gaussian for each value of $t$. 
As an example,  Figure \ref{fone}  presents $P(g)$ for systems with $t=0.05$ and $t=1$.
The inset of the figure \ref{fone} shows that the mean conductance is always larger than 1. 
The variance var $g=\langle g^2\rangle-\langle g\rangle^2$  increases
as $t$ decreases. The system size dependence of the mean conductance is presented 
in Fig. \ref{ftwo} for both  the Q1D and the square samples.
For $t\ge t_c\approx 0.2$,
the mean conductance decreases as  $\langle g\rangle\propto L_y/L_x$ in the Q1D case and
is almost system-size independent for  the squares. This confirms that these  systems are
in the diffusive regime.
For a stronger anisotropy, $t<t_c$, the diffusive regime takes place for  much smaller
systems ($30\le L\le 50$ for $t=0.1$). 

The analytical expression for var $g$ derived by  Stone et al.
\cite{stone} 
states that  the anisotropy influences the variance of the conductance only in the
combination with the size of the system:
\be\label{fluct}
{\rm var}~ g = f\left(\frac{L_x}{L_y}\sqrt{t}\right).
\ee
Fig. \ref{fthree}  shows  that for $L_x/L_y\sqrt{t}>2$ and $t\ge t_c$, var $g$
reaches the universal value = 2/15. 
For the squares, var $g$ is independent on the system size for $t>0.1$.

\smallskip

The increase of var $g$ for $t< 1$ is in a qualitative agreement with the universal
relation for the conductance fluctuation
\cite{stone,MK}.  
\be\label{vg}
{\rm var} g\sim \beta^{-1}
\ee
provided that $\beta<1$ when $t<1$. It seems straightforward to 
compare numerical data for var $g$ with Eq. \ref{vg} and
calculate $\beta=\beta(t)$.  However, detailed numerical analyses have shown that
relation (\ref{vg})
underestimates numerical data if the disorder is weak.
\cite{japonci,RMS}
Therefore we prefer to estimate $\beta(t)$ 
from the statistics of the parameters $z$ for anisotropic  
square samples. 
The quantity of interest is the probability distribution $P(s)$ of
normalized differences \cite{pozn}
\be\label{nd}
s=(z_{i+1}-z_i)/\langle z_{i+1}-z_i\rangle 
\ee
Fig. \ref{ffour}  shows the non-analytical behavior  
\be\label{smalls}
P(s)\sim s^{\beta(t)}\quad{\rm for}\quad s<<1.
\ee
The  $t$-dependent exponent $\beta=\beta(t)$ (inset of fig. \ref{ffour}) is calculated from the 
logarithmic fit 
\be\label{betafit}
\log P(\log s) = [1+\beta(t)]~\log s
\ee
Relation (\ref{smalls}) with $\beta=1,2$ or 4 is well known from the RMT. The small $s$ behavior
is determined by the  symmetry of the model. Here, however,  
$\beta$ is given by  statistical properties of the system.
In the limit $t\to 0$  
our system dissociates to a set of independent
chains, each of them is characterized by its  $z$.
Hence  $z$  are statistically independent variables and the
distribution of  their difference
should be Poissonian: $P(\log s)\propto \log s$ and $\beta(t=0)=0$.
The exponent  $\beta(t)$ increases as $t$ increases and the distribution
$P(s)$  converges to the Wigner surmise (WS) as $t\to 1$: $P(\log s)\propto 2\log s$ as expected.

Contrary  to $P(s)$, the distribution $P(z_1)$ of the smallest $z_1$ is  similar
to WS for any $t$. This is consistent with RMT 
which states that the distribution of $z_1$ is $\beta$-independent.
\cite{pichardnato}

To test the applicability of  RMT with non-integer $\beta$   
to anisotropic systems,  we studied the  
spectra of $z$ for squares (Fig. \ref{ffive}).
Assuming that the distribution $P(z)$ has the form (\ref{rmtp},\ref{rmt}),
we can find the most probable values $\tilde{z}$
of  the parameters $z$  from the
system of nonlinear equations
\be\label{rov}\partial{\cal H}/\partial z_i\Big|_{z_i=\tilde{z}_i} =0.\ee

In the limit of small $z$ 
the "interaction" and "Jacobian" terms in (\ref{rmt}) can be approximated as
\cite{pichardnato}.
\be\label{interaction}
u(z_i,z_j)=\log|z_i^2-z_j^2|\quad {\rm and}\quad
J(z)=\log z.
\ee
System (\ref{rov}) is then exactly solvable
\cite{Muttalib}.  After some algebra we find
\cite{JPCM}
\be\label{spectrum}
{\tilde z}_i\sim j_\alpha(i), \quad\quad \alpha=\frac{1}{\beta}-1
\ee
where $j_\alpha(i)$ is the $i$th zero of the Bessel function $J_\alpha$.
From (\ref{spectrum}) we can express the ratio
\be\label{ratio}
\frac{\tilde{z}_{i+1}}{\tilde{z}_i}=\frac{j_\alpha(i+1)}{j_\alpha(i)}
\ee
which depends only on $\beta$.\cite{p2} 

The derivation of (\ref{ratio}) holds for any value of $\beta$.
Using the values of $\beta(t)$ presented in Fig. \ref{ffour}
we calculated
the ratio $\tilde{z}_{i+1}/\tilde{z}_i$ and
compared it with  numerical data. As it is shown in  inset of Fig. \ref{ffive},
the agreement is very good for  $\alpha<2$, which corresponds to $t\ge t_c=0.2$. 

\smallskip

In Fig. \ref{fsix} we show that the exponent $\beta$ depends also on the 
shape of the system.  $\beta$ converges toward 1 as the length $L_x$ increases.
This is consistent with Ref. \cite{MK}. 
For  $t=0.2$ we found that the distribution  $P(s)$ is  WS  
for $L_x/L_y\approx 8$. For this system length the system is still in the metallic regime:
the mean conductance $\langle g\rangle\sim 1$, and the RMT with $\beta=1$
is applicable to describe its properties.
Of course, further increase of  the  system length causes a decrease of
the mean conductance and $P(s)$ becomes Gaussian. A qualitatively similar 
behavior can be found for any $t>0.2$.

For $t=0.05$, $P(s)$ reaches  WS only 
for $L_x/L_y\approx 36$. However, the coincidence of $P(s)$ with WS does not indicate
the metallic behavior. 
The  mean conductance is $\langle g\rangle\sim 10^{-2}$.
Thus we have an  interesting paradox: the strongly anisotropic system exhibits
the  metallic behavior 
with  a distribution $P(s)$ very close to Poissonian distribution.
By increasing  the
system length we obtain an insulating regime in which $P(s)$ becomes  WS.

\smallskip

In conclusion, we have presented numerical data for the strongly anisotropic weakly disordered systems.  
For $t\ge t_c\approx 0.2$ we found the metallic behavior with the mean 
conductance independent on the system size (for size $L<100$). The distribution of the
conductance is Gaussian. 
We found that the anisotropy causes the increase of the var $g$.
We analyzed also the spectrum of
the parameters $z$. We found that the shape of the distribution 
$P(s)$ of the normalized difference 
$s$ (\ref{nd}) depends on the anisotropy.
We interpret these results  by the random matrix theory
in which the "symmetry parameter" $\beta$ depends on the anisotropy and can possess any
positive value. From such RMT we derived the  analytical formula for the 
spectrum of  $z$  which agrees very well with numerical data.

The assumption that $\beta$ could be non-integer corresponds  with  the theoretical prediction of
Muttalib and Klauder.
\cite{MK} In their theory,  DMPK equation can be  
generalized to the  description of shorter
systems.  The parameter  $\beta$ becomes then  a function of  
mutual correlations of eigenvalues and eigenvectors
of the transfer matrix.  In agreement with Ref. \cite{MK},
we found that $\beta$ depends on the length  of the system and
converges to 1 when  the system length increases.
 
Another, more formal interpretation of RMT with non-integer $\beta$ 
is based on the Coulomb gas analogy: \cite{pichard}
the probability distribution (\ref{rmtp}) is formally 
equivalent to the statistical weight of the classical system 
of charged interacting particles 
in one dimension. $z$ determines  position of the  particles 
which interact via interaction $\beta u(z_i,z_j)$.
The parameter $\beta$ represents   the strength of  interaction.
The anisotropy parameter $t$ tunes this  interaction.
The limit $t=0$ represents the system of non-interacting particles
with $\beta=0$. 
This  effect of weak interaction must not be confused with the decrease of the
interaction which appears in the isotropic Q1D system. 
In the last phenomena the confining one-particle potential $kz^2/2$ becomes weaker
as the length of the system increases. This enables an increase of the  
mutual distance  $z_{i+1}-z_i$ between particles. The effect of the interaction
(which is a function of the particle distance) is therefore less important
than in the metallic (short) systems. This does not affect the value
of the interaction constant $\beta$.

The critical value of anisotropy $t_c=0.2$ appears frequently  throughout the paper.
For stronger anisotropy (smaller $t$), no diffusive regime exists.
The actual value of $t_c$ is, of course, determined by our choice
of the strength of disorder and is expected  to be smaller for 
smaller $W$.

We suppose that the anisotropic model  discussed in this paper represents the  
physical system to which the generalized DMPK equation of Muttalib and Klauder
\cite{MK}  can be applied.

\smallskip

\baselineskip=10pt{\small \noindent
This work was supported by Slovak Grand Agency VEGA, Grant n. 2/7174/20. Numerical data were partially collected using the computer Origin 2000 in the Computer Center of the Slovak Academy of Sciences.

\newpage

\begin{figure}
\noindent\epsfig{file=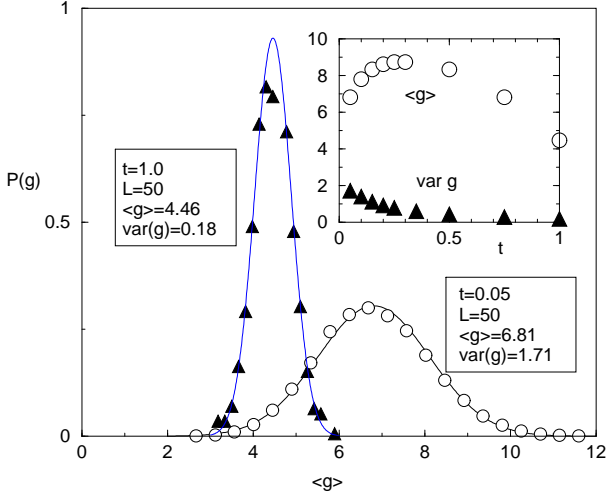,width=8cm}
\vspace*{2mm}
\caption{The distribution of the conductance for $t=0.05$ and $t=1$.
Inset: The $t$-dependence of $\langle g\rangle$ and var $g$ for squares
$50\times 50$.}\label{fone}
\end{figure}

\begin{figure}
\noindent\epsfig{file=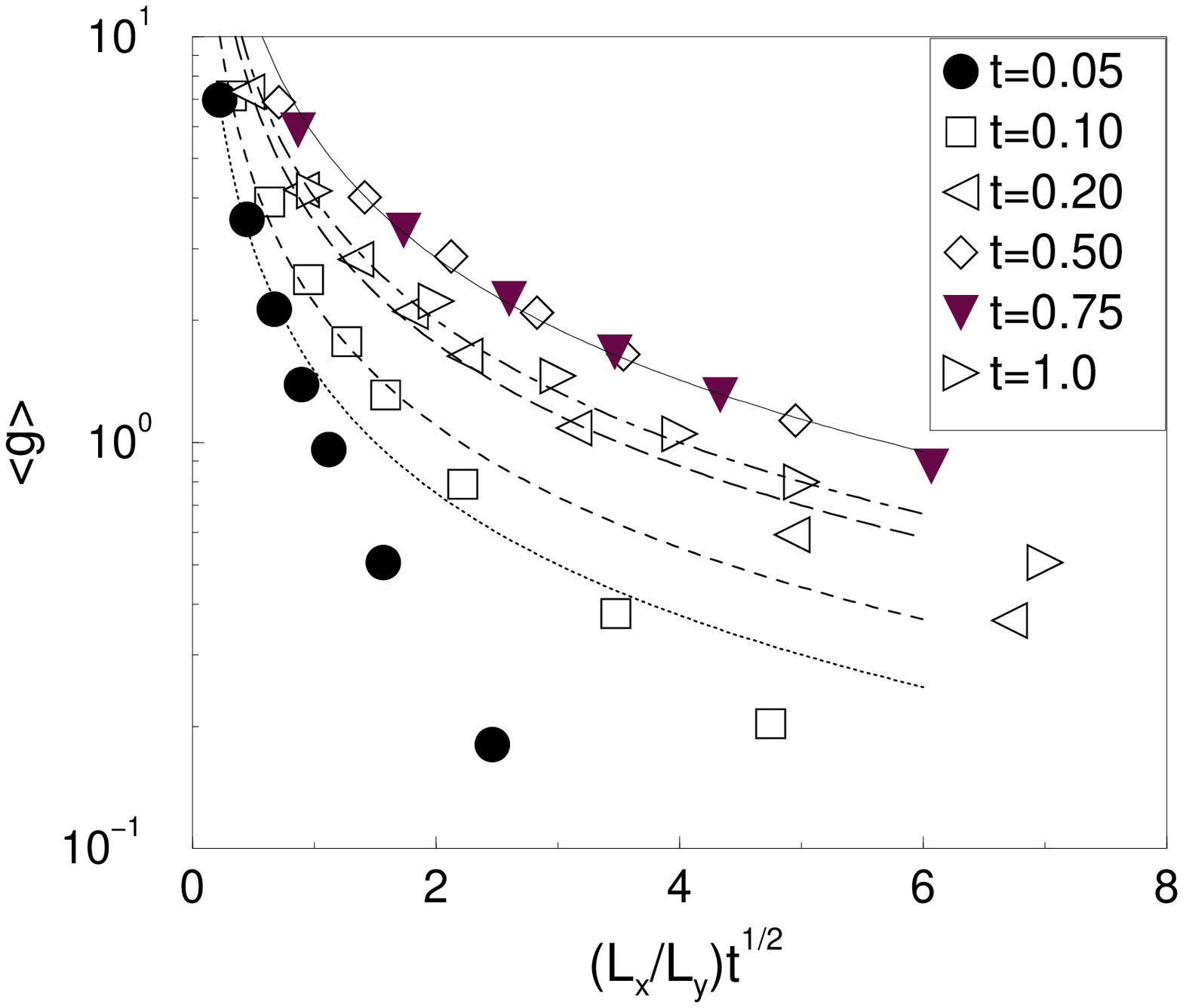,width=4cm}~~\epsfig{file=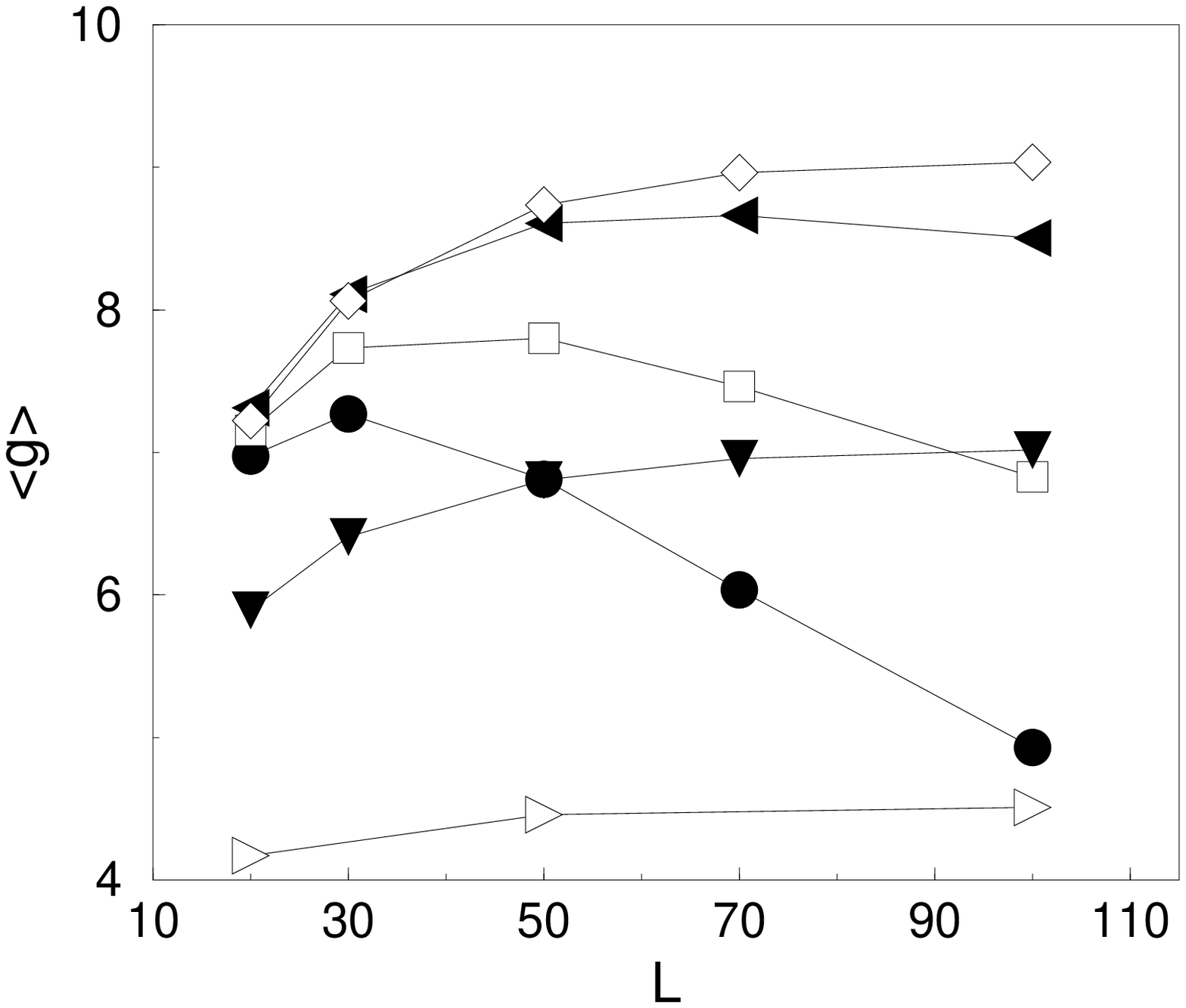,width=4cm}
\vspace*{2mm}
\caption{System size dependence of the mean conductance  $\langle g\rangle$.
Left: for the quasi-one dimensional systems $L_x\times L_y$
The width of the system is $L_y=10$.   
Lines represent 
the  relation $\langle g\rangle\sim a\times L_x/L_y$ which is characteristic for the
diffusive regime. 
Right:  for squares $L\times L$. Increase of the mean conductance for small $L$ 
indicates a ballistic regime, decrease for large $L$ is due to the crossover to the localized
regime.  }\label{ftwo}
\end{figure}

\begin{figure}
\noindent\epsfig{file=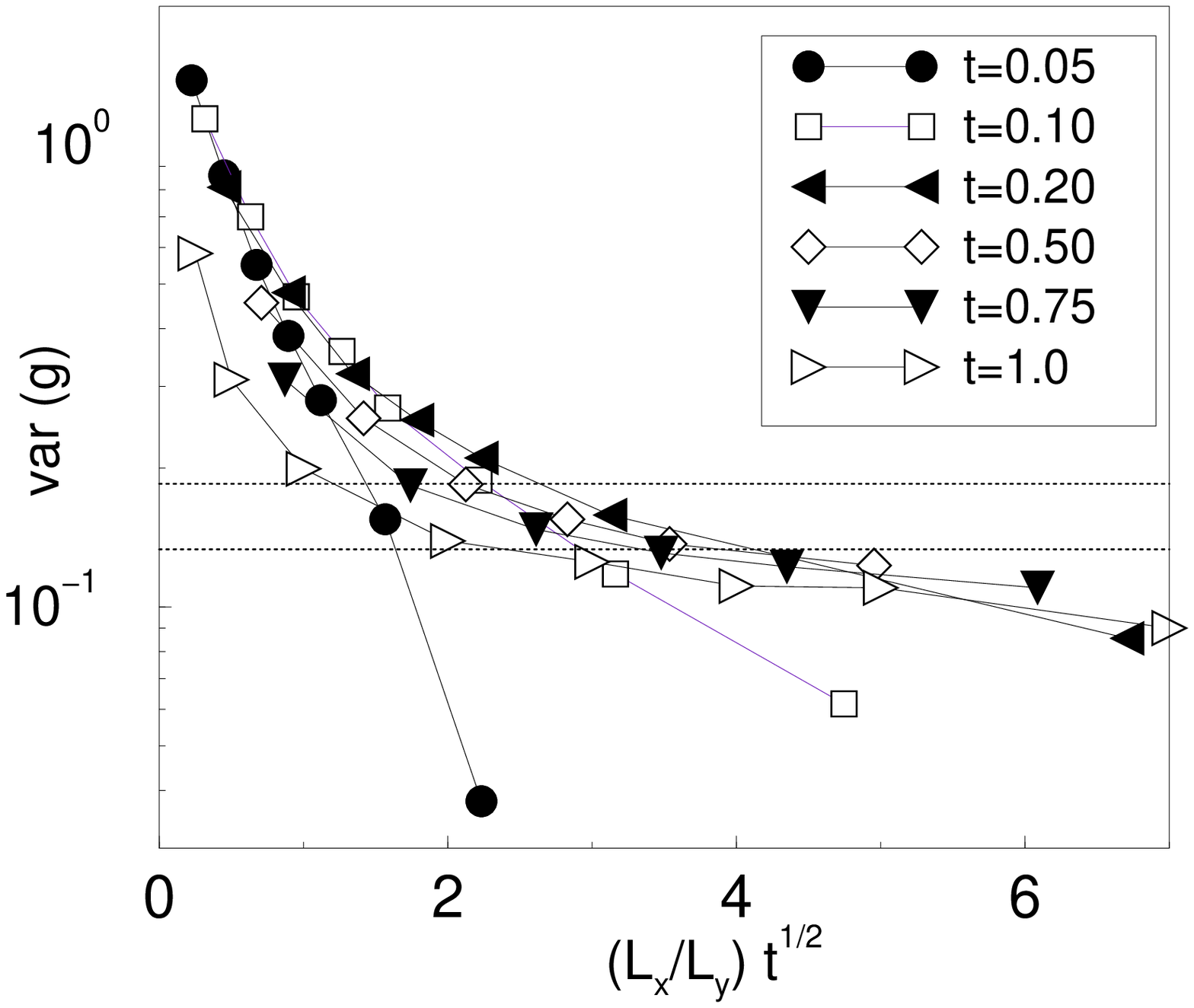,width=4cm}~~\epsfig{file=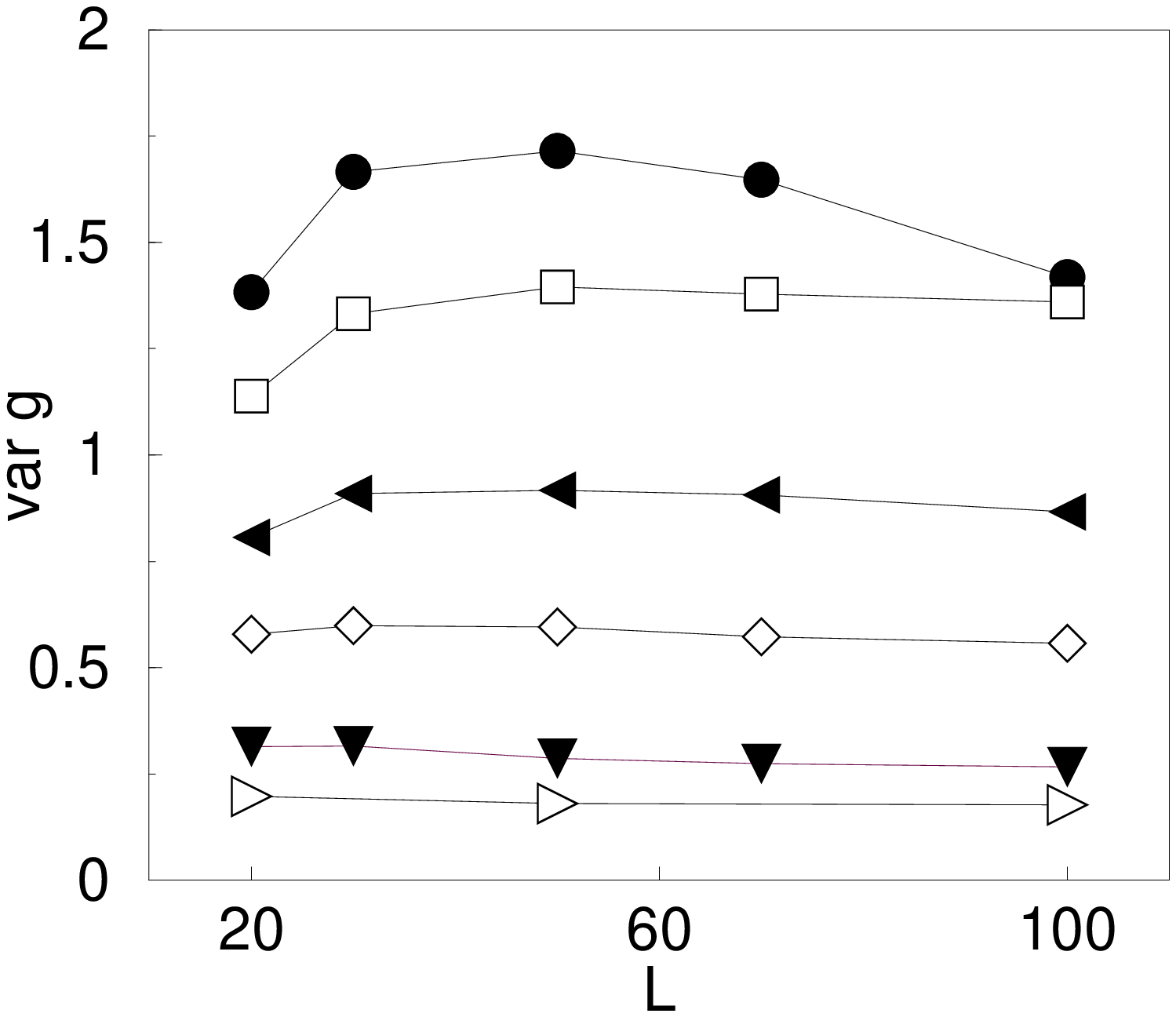,width=4cm}
\vspace*{2mm}
\caption{Left: var $g$ as a function of 
the system length $L_x$. The width of the system $L_y=10$. Dotted lines show universal values of var $g$ for
squares (0.185) and Q1D systems (0.133) \cite{stone}. 
The increase of the system length causes
the transition to localized regime with decrease of  var $g$.
For $t=1$, also data for shorter systems
$40\times 10$ and $20\times 10$ are present to shown an increase of the
conductance fluctuations.
Right:  var $g$ as a function of the system size for squares $L\times L$. 
}\label{fthree}
\end{figure}

\begin{figure}
\noindent\epsfig{file=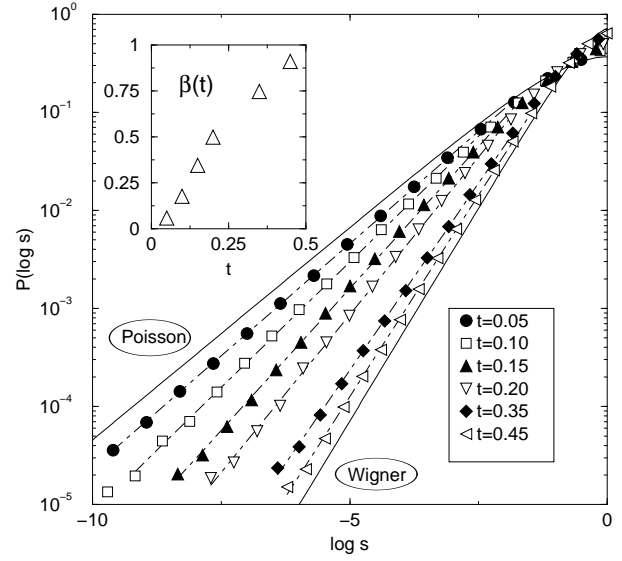,width=8cm}
\vspace*{2mm}
\caption{Probability density  $P(\log s)$ for different anisotropy $t$ of the system.   $s$ is  the
(normalized) difference $z_{i+1}-z_i$. 
Solid lines are 
Wigner surmise $W_1(s)=\frac{\pi}{2}s\exp -\frac{\pi}{4}s^2$ and Poisson
distribution $e^{-s}$. 
The size of the samples $L_x=L_y=50$. 
Statistical ensembles of  $N_{\rm stat}=10^5$
samples have  been considered. 
Dot-dashed lines represent fits (\ref{betafit}).
Inset:  $t$- dependence of exponent $\beta$.
}\label{ffour} 
\end{figure}

\begin{figure}
\noindent\epsfig{file=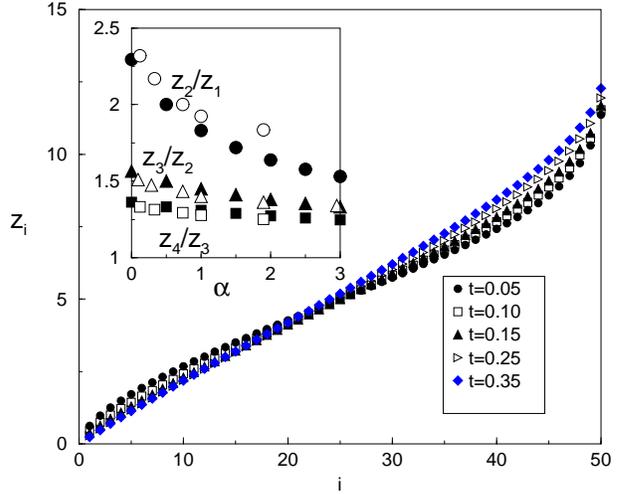,width=8cm}
\vspace*{2mm}
\caption{Spectrum of $z$s for small  anisotropy parameters. Note the common crossing point of spectra for different $t$. Inset: comparison of
ratios of the most probable values,  $z_{i+1}/z_i$ for $i=1,2,3$ (open symbols) with theoretical prediction $j_\alpha(i+1)/j_\alpha(i)$ full symbols). Parameter $\alpha=\beta^{-1}-1$ (\ref{spectrum}).}\label{ffive}
\end{figure}

\begin{figure}
\noindent\epsfig{file=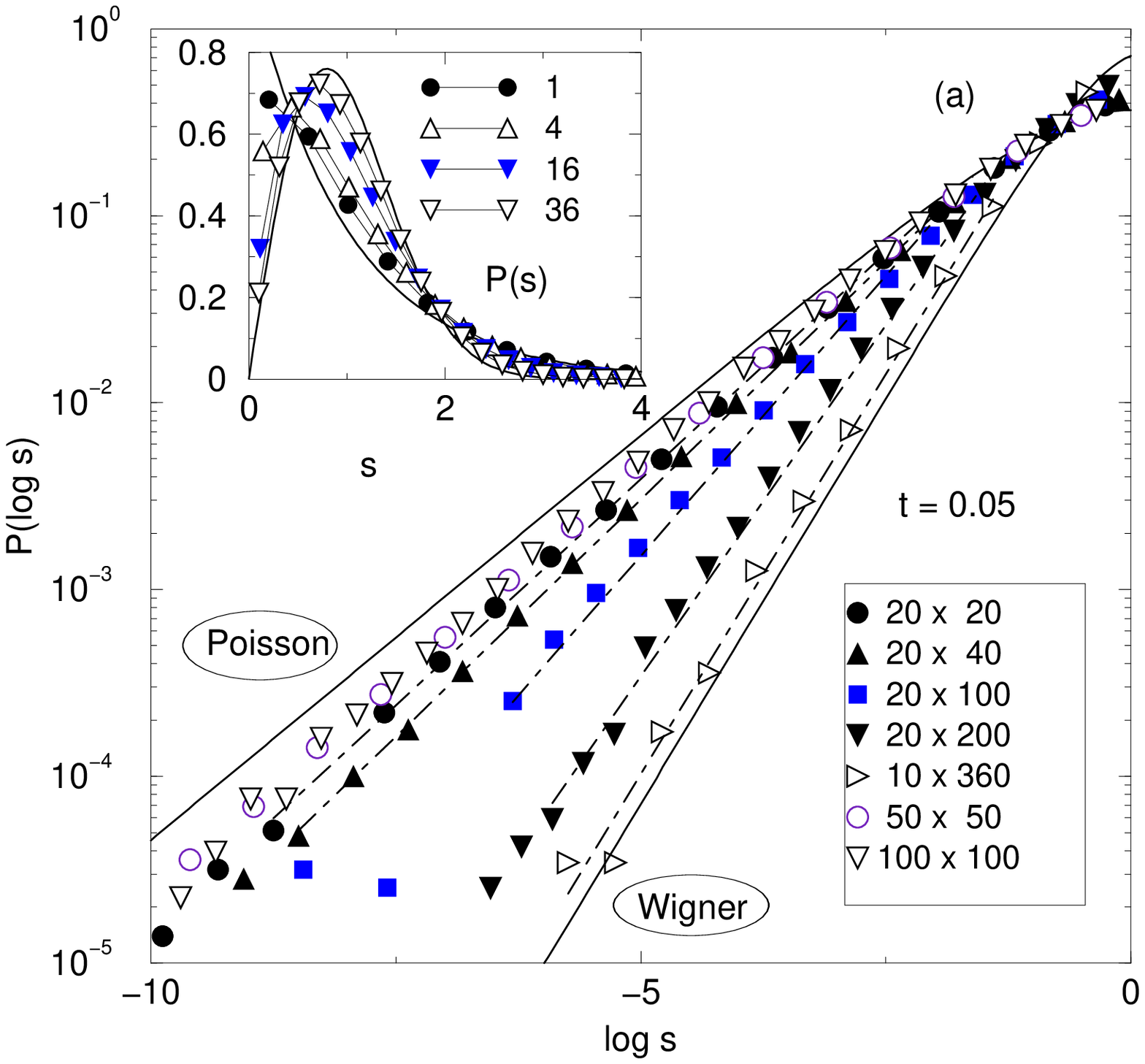,width=7cm}
\vspace*{2mm}
\noindent\epsfig{file=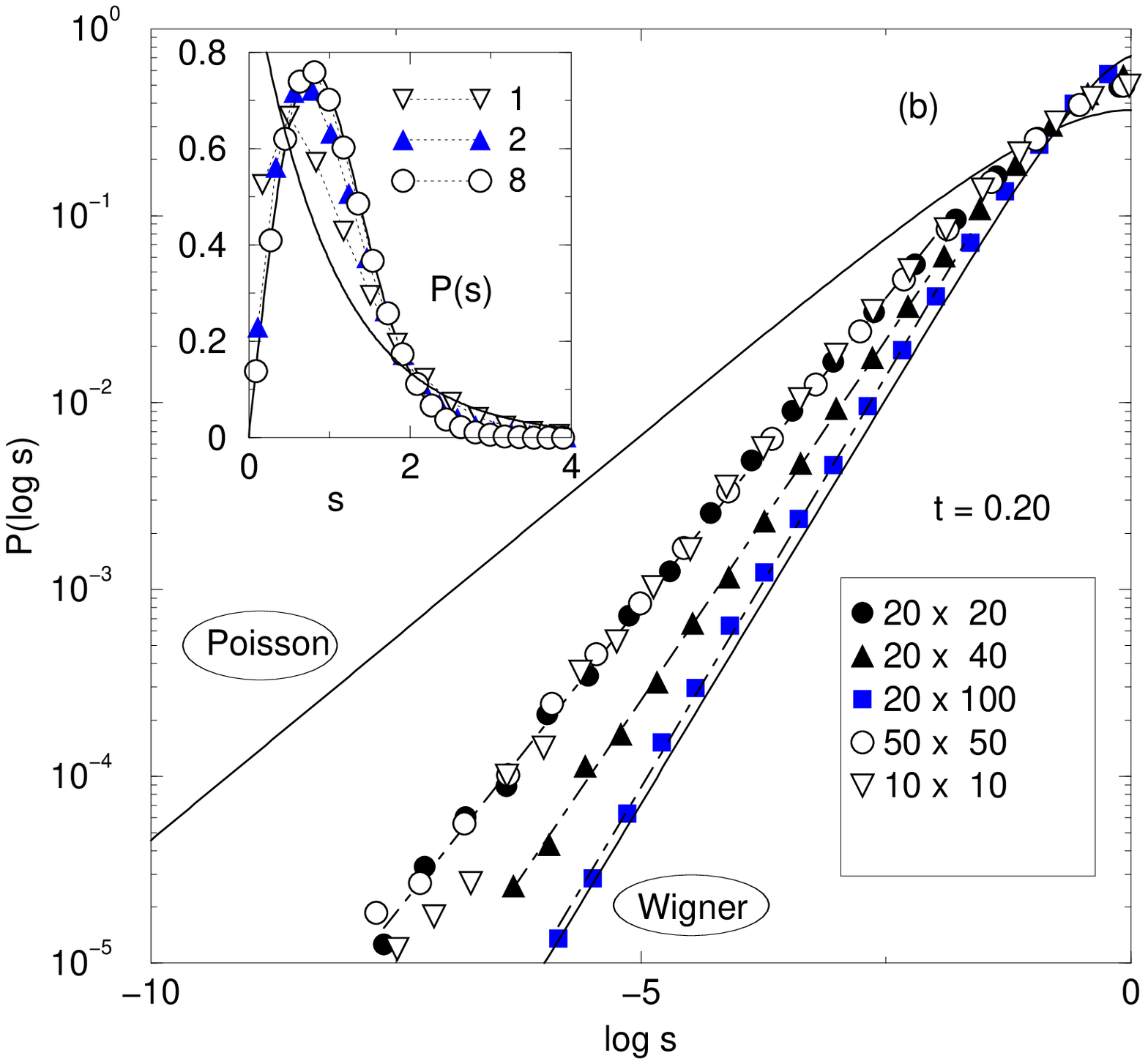,width=7cm}
\vspace*{2mm}
\caption{Change of $P(\log s)$ with the length of the system.
For  a longer system, the exponent $\beta$ increases and $P(s)$ converges to Wigner
surmise. Insets show $P(s)$ in linear scale for some ratio $L_x/L_y$.
 (a) $t=0.05$: $P(s)$ achieves Wigner surmises for $L_x/L_y>36$. Comparison
with Fig. 2 shows that the conductance of such long system is
small. (b) $t=0.2$. $P(s)$ has a form of Wigner surmises already for
$L_x/L_y=8$, when the system is still in a metallic state, $\langle g\rangle\sim 1 $.
We present also $P(s)$ for square samples of various size to show that
exponent $\beta$ is system-size independent although $\langle g\rangle$ is not
constant (see figure 2).}\label{fsix}
\end{figure}


\begin{thebibliography}{99}
\bibitem{p2} In \cite{JPCM} we applied   relation (\ref{ratio}) to orthogonal ($\beta=1$)
and symplectic $\beta=4$) systems to describe the exact form of the
spectrum of $z$. In particular, formula (\ref{ratio}) predicts that  $z_2/z_1\approx 2.3$ for the
orthogonal and $\approx 4.94$ for the symplectic ensembles. This agrees with numerical data.
\end{thebibliography}
\end{document}